%% file: OptimizeRegPower.tex
\documentclass[format=acmsmall, review=false, screen=true,rgb]{acmart}

\renewcommand\footnotetextcopyrightpermission[1]{} 
\usepackage{booktabs}   
\usepackage{subcaption} 
                        
\usepackage{lmodern}
\usepackage{xspace}
\usepackage{etex}
\usepackage{booktabs} 
\usepackage{amsmath}
\usepackage{graphicx}
\usepackage{caption}
\usepackage[margin=1pt,font=footnotesize,textfont=bf,labelfont=bf]{subcaption}
\usepackage{listings}
\usepackage[margin=1pt,font=footnotesize,textfont=bf,labelfont=bf]{caption}
\usepackage{ftnxtra}
\usepackage{threeparttable}
\usepackage{balance}
\usepackage{courier}
\usepackage[pdf]{pstricks}
\usepackage{pst-rel-points}
\usepackage{pst-node}
\usepackage{algpseudocode}
\usepackage{pslatex}
\usepackage{listings}
\usepackage{comment}
\usepackage{wrapfig}
\usepackage{tabularx}
\usepackage[inline]{enumitem}
\setlist{nolistsep,leftmargin=*}
\usepackage{multirow}
\usepackage{url}
\usepackage{breakurl} 
\usepackage[ruled]{algorithm2e} 
\usepackage{subcaption}
\usepackage{footmisc}
\usepackage{xcolor,soul}

\SetAlFnt{\small}
\SetAlCapFnt{\small}
\SetAlCapNameFnt{\small}
\SetAlCapHSkip{0pt}
\IncMargin{-\parindent}

\setcopyright{none}

\include{Sections/defines}

\renewcommand{\textrightarrow}{$\rightarrow$}

\newcommand{\ourtool}{\mbox{\bf GR{\sc e}E{\sc ne}R}\xspace}

\begin{document}

\title{\ourtool: A Tool for Improving Energy Efficiency of Register Files}
  
\author{Vishwesh Jatala}
\affiliation{%
  \institution{Indian Institute of Technology, Kanpur}
  \department{Department of CSE}
  \city{Kanpur}
  \state{Uttar Pradesh}
  \postcode{208016}
  \country{India}  
}
\email{vjatala@cse.iitk.ac.in}
\author{Jayvant Anantpur}
\affiliation{%
  \institution{Mentor Graphics India Pvt Ltd}
  \department{}
  \city{Bangalore}
  \state{Karnataka}
  \postcode{560012}
  \country{India}
}
\email{jayvant.anantpur@gmail.com}
\author{Amey Karkare}
\affiliation{%
\institution{Indian Institute of Technology, Kanpur}
  \department{Department of CSE}
  \city{Kanpur}
  \state{Uttar Pradesh}
  \postcode{208016}
  \country{India}
} 
\email{karkare@cse.iitk.ac.in}

\input{Sections/0-Abstract}



\maketitle

\renewcommand{\shortauthors}{V. Jatala et al.}

\input{Sections/01-Introduction}

\input{Sections/02-Background}

\input{Sections/04-Implementation}

\input{Sections/07-EvaluationMethodology}

\input{Sections/08-ExperimentalAnalysis}

\input{Sections/09-RelatedWork}

\input{Sections/10-Conclusion}

\bibliographystyle{ACM-Reference-Format}
\bibliography{OptimizeRegPower}

\end{document}

%% file: Sections/defines.tex
 \newcounter{lcount}

\newcommand{\cmt}[1]{}

\newcommand{\moveup}{\vskip -4mm}

\newcommand{\inc}[1]{\ensuremath{{\sf INC}(#1)}}

%% file: Sections/0-Abstract.tex
\begin{abstract}

Graphics Processing Units (GPUs) maintain a large register file to increase the thread level parallelism (TLP). To increase the TLP further, recent GPUs have increased the number of on-chip registers in every generation. However, with the increase in the register file size, the leakage power increases. Also, with the technology advances, the leakage power component has increased and has become an important consideration for the manufacturing process. The leakage power of a register file can be reduced by turning infrequently used registers into low power (drowsy or off) state after accessing them. A major challenge in doing so is the lack of runtime register access information. 


This paper proposes \ourtool ({\sc Gpu} REgister file ENErgy
Reducer): a system to minimize leakage energy of the register
file of GPUs. \ourtool employs a compile-time analysis to
estimate the run-time register access information. The
result of the analysis is used to determine the power state of
the registers (ON, SLEEP, or OFF) after each instruction. We
propose a power optimized assembly instruction set that allows
\ourtool to encode the power state of the registers in the
executable itself. The modified assembly, along with a run-time
optimization to update the power state of a register during
execution, results in significant power reduction.

We implemented \ourtool in GPGPU-Sim simulator, and used
GPUWattch framework to measure the register file's leakage
power. Evaluation of \ourtool on 21 kernels from CUDASDK,
GPGPU-SIM, Parboil, and Rodinia benchmarks suites shows an
average reduction of register leakage energy by 69.04\% and
maximum reduction of 87.95\%  with a negligible number of 
simulation cycles overhead (0.53\% on average).

\end{abstract}

%% file: Sections/01-Introduction.tex
\section{Introduction}
\label{sec:intro}

Graphics Processing Unit (GPU) achieves high throughput by utilizing thread level parallelism (TLP). Typically, GPUs maintain a large register file in each streaming multiprocessor (SM) to improve the TLP. GPUs allow a large number of resident threads~\cite{kepler} in each SM, and the resident threads can store their thread context in the register file, which facilitates faster context switching of the threads. The threads that are launched in each SM are grouped into sets of 32 threads (called warps), and they execute the instructions in a single instruction, multiple threaded (SIMT) manner. To keep improving the TLP of the GPUs, GPU architects increase the maximum number of resident threads and  register file sizes in  every generation.  For instance, NVIDIA Fermi GF100 has 128KB register file and allows up to 1536 resident threads, while NVIDIA Kepler GK110 has 256KB register file  and allows maximum 2048 resident threads~\cite{kepler}. 

Earlier studies~\cite{GPUWattch,RegVirtual} show that register files in GPUs consume around 15\% of the total power. With the technology advances, the leakage power component has increased and has become an important consideration for the manufacturing process~\cite{Moore}. Moreover, registers in a GPU continue to dissipate leakage power throughout the entire execution of its warp even when they are not accessed by the warp.

\begin{figure}[t]
  \centering
  \includegraphics[scale=0.27]{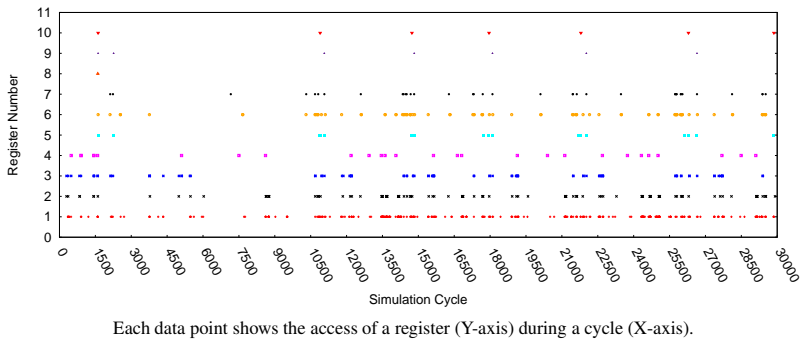}              
  \vskip -1mm
  {\scriptsize Each data point shows the access of a register (Y-axis) during a cycle (X-axis).}
  \caption{Register Access Pattern for MUM~\cite{GPGPU-Sim}}   
  \label{fig:MUM}
  \vskip -3mm
\end{figure}  

\vspace{-3mm}
\subsection{Motivation}
\label{sec:motivation} 

\begin{figure}
\centering
\includegraphics[scale=0.38]{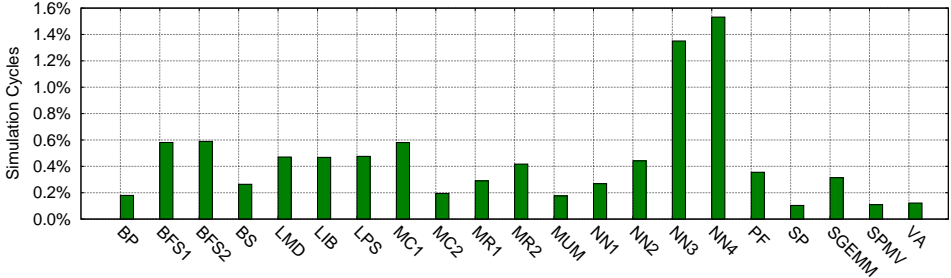}
\vskip -5mm
\caption{Percentage of Simulation Cycles Spent by a Register (Averaged Over all the Registers)}
\label{fig:perc_reg_access}
\vskip -5mm
\end{figure}

To understand the severity of leakage power dissipation by
register file, consider Figure~\ref{fig:MUM} which shows the
access patterns of some registers of warp 0 during the execution
of \emph{MUM} application (The experimental methodology has been discussed in Section~\ref{sec:experimentanalysis}). 
We use the access patterns of the registers of a
single warp as a representative since all the warps of a kernel
typically show similar behavior during execution~\cite{pilot}. We
make the following observations:
\begin{itemize}
\item Register 10 is accessed very infrequently---it is accessed for
  only 7 cycles during the complete execution
  (life time) of the warp (29614 cycles).
\item Register 1 is the most frequently accessed register during the warp
  execution. However, it is accessed for only 330 cycles
  ($\sim$ 1.11\%) during the life time of the warp.
\end{itemize}
This shows that registers are accessed for a very short duration
during the warp life time. However, they continue to dissipate
leakage power for the entire life time of the
warp.  Figure~\ref{fig:perc_reg_access} shows that the behavior
is not specific to \emph{MUM}, but is seen across a wide range of
applications. The figure shows the percentage of simulation
cycles spent in register accesses (averaged over all the
registers in all the warps) for several applications.  We observe
that registers on an average spend $<2\%$ of the simulation
cycles during the warp execution while leaking power during the
entire execution.

One solution~\cite{WarpedRegFile} to reduce the leakage power of
the registers is by putting the registers into drowsy or
SLEEP\footnote{Drowsy~\cite{WarpedRegFile,Drowsy} and
  SLEEP~\cite{Mcpat} states refer to the same low power data
  preserving states. In this paper, we use the term SLEEP.} state
{\em immediately} after the registers of an instruction are
accessed. However, this can have run-time overhead whenever there
are frequent wake up signals to the sleeping register. Consider
Figure~\ref{fig:MUM} again:

\begin{itemize}
\item Putting register 10 to SLEEP state immediately after its
  accesses saves significant power due to the gaps of
  several thousands of cycles between consecutive accesses.
\item In contrast, register 1 is accessed very frequently.  If it
  is put to SLEEP after every access, it will have a high overhead
  of wake up signals.
\item The access pattern of register 7 changes during the warp
  execution. It is accessed frequently for some duration (for
  example, between cycles 10500--11250), and not accessed frequently for
  other duration (between cycles 3000--7500). To optimize energy
  as well as run-time, the register needs to be kept ON whenever
  it is frequently accessed, and put to SLEEP otherwise.
\item The last access to register 8 is at cycle 1602. The
  register can be turned OFF after its last access to
  save more power.
\end{itemize}

In summary, the knowledge of registers' access patterns helps improve
energy efficiency without impacting the run-time adversely. Our proposed solution \ourtool statically estimates the run-time usage patterns of registers to reduce GPU register file leakage power.

\subsection{Contributions}

\ourtool uses a compile-time analysis to determine the power
state of the registers (OFF, SLEEP, or ON) for
each instruction by estimating the register usage
information. Further, it transforms an input assembly language by
encoding the power state information at each instruction to make
it energy efficient. The static analysis makes safe
approximations while computing power state of the registers,
therefore, the choice of the state can be suboptimal at
run-time. Hence, to improve the accuracy and energy efficiency,
it provides a run-time optimization that dynamically corrects the
power state of registers of each instruction. We make the following contributions:
\begin{enumerate}
\item We introduce a new instruction format that supports the
  power states for the instruction registers (Section~\ref{sec:encoding}).
  We propose a compile-time analysis that determines the
  power state of the registers at each program point and
  transforms an input assembly language into a power optimized
  assembly language (Section~\ref{sec:companalysis} and \ref{sec:encoding}).
\item We propose a run-time optimization to reduce the penalty of
  suboptimal (but safe) choices made by static analysis (Section~\ref{sec:runtime}).
\item We implemented the proposed compile-time and run-time
  optimizations using GPGPU-Sim simulator~\cite{GPGPUSIM}. 
  We integrated GPUWattch~\cite{GPUWattch} with CACT-P~\cite{Cactip} version to 
  enable power saving mechanism (Section~\ref{sec:experimentanalysis}). 
\item   We evaluated our implementation on wide range of kernels from
  different benchmark suites: CUDASDK~\cite{CUDA-SDK},  
  GPGPU-SIM~\cite{GPGPU-Sim}, Parboil~\cite{parboil}, 
  and Rodinia~\cite{rodinia}. We observe a reduction in the register leakage energy by an  average of 69.04\% and maximum of 87.95\% (Section~\ref{sec:experimentanalysis}). 
  
\end{enumerate}

In the paper, Section~\ref{sec:background} briefs the background
required for \ourtool, while the system itself is described in
Section~\ref{sec:implementation}. 
Section~\ref{sec:experimentanalysis} and \ref{sec:results}
give the experimental evaluation. Section~\ref{sec:relatedwork}
describes related work, and Section~\ref{sec:conclusion}
concludes the paper.

%% file: Sections/02-Background.tex
\section{Background} \label{sec:background}
GPUs consist of a set of streaming multiprocessors (SMs). Each SM contains a large number of execution units such as ALUs, SPs, SFUs, and Load/Store units. GPUs achieve high throughput because they can hide long memory execution latencies with massive thread level parallelism. Each SM has a large register file, which allows the resident threads to maintain their contexts, and hence can have faster context switching. To reduce the access latency, the register file is divided into multiple banks. The registers from different banks can be accessed in parallel.  A bank conflict occurs whenever multiple registers need to be accessed from the same bank, and these need to be  accessed in serial.  In GPGPU-Sim simulator~\cite{GPGPUSIM} the requests for instruction registers are stored in a collector unit. When all the operands of the instruction are ready, it can proceed to the execution stage.

NVIDIA provides a programming language CUDA~\cite{CUDA} to parallelize applications on GPU. The portion of the code which is to be parallelized is specified using a special function called kernel. A kernel is invoked with the number of thread blocks and the number of threads in each thread block as parameters. A program written in CUDA can be compiled using \emph{nvcc} compiler. The compiler translates the program into an intermediate representation (PTX), which is finally translated to an executable code. NVIDIA provides tools such as \emph{cuobjdump} to disassemble the executable into SASS assembly language. GPGPU-Sim converts SASS code to PTXPlus code for simulation.

GPUWattch~\cite{GPUWattch} framework uses the simulation statistics of GPGPU-Sim to measure the power of each component in the GPUs. The framework is built on McPAT~\cite{Mcpat}, which internally uses CACTI~\cite{Cacti}. McPAT models the register files as memory arrays to measure the register power. CACTI divides memory arrays into set of banks, which are finally divided into subarrays (collection of memory cells).

\ourtool optimizes the PTXPlus code to make it energy efficient
by reducing the leakage power of the register files. Our
experiments use GPUWattch framework to measure the leakage power.

%% file: Sections/04-Implementation.tex
\newcommand{\fgentry}{{\sf Entry}}
\newcommand{\fgexit}{{\sf Exit}}
\newcommand{\IN}[1]{\ensuremath{{\sf IN}(#1)}}
\newcommand{\OUT}[1]{\ensuremath{{\sf OUT}(#1)}}
\newcommand{\Pred}[1]{\ensuremath{{\sf PRED}(#1)}}
\newcommand{\Succ}[1]{\ensuremath{{\sf SUCC}(#1)}}
\newcommand{\islive}[2]{\ensuremath{{\sf isLive}(#1, #2)}}
\newcommand{\dist}[2]{\ensuremath{{\sf Dist}(#1, #2)}}
\newcommand{\sleepoff}[2]{\ensuremath{{\sf SleepOff}(#1, #2)}}
\newcommand{\power}[2]{\ensuremath{{\sf Power}(#1, #2)}}

\section{\ourtool} \label{sec:implementation}
To understand the working of \ourtool, we need to understand
the different access patterns of a register and their effect
on the wake up penalty incurred. Let $W$ (threshold) denotes the minimum number of
program instructions that are required to offset the wake-up
penalty incurred when a register state is switched from OFF or
SLEEP state to ON state.
%
Consider a program that accesses some register $R$ in a statement
$S$ during execution. The future accesses of $R$ in this
execution govern its power state. The following scenarios exist:
\begin{enumerate}
\item The next access (either read or write) to $R$ is by an
  instruction $S'$ and there are no more than  $W$ instructions
  between $S$ and $S'$.  In this case, since the two accesses to
  $R$ are very close, it should be kept ON to avoid any wake-up
  penalty associated with SLEEP or OFF state.
\item The next access to $R$ is a read access by an instruction
  $S'$ and there are more than $W$ instructions between $S$ and
  $S'$. In this case, since the value stored in $R$ is used by
  $S'$, we can not switch $R$ to OFF state as it will cause the loss of
  its value. However, we can put $R$ in SLEEP state.
\item The next access to $R$ is a write access by an instruction
  $S'$ and there are more than $W$ instructions between $S$ and
  $S'$. In this case, since the value stored in $R$ is being
  overwritten by $S'$, we can put $R$ in OFF state.
\item There is no further access to $R$ in the program. In this
  case also, register $R$ can be safely turned OFF.
\end{enumerate}
We now describe the compiler analysis used by \ourtool to capture
these scenarios.

\subsection{Compiler Analysis} \label{sec:companalysis} 
To compute power state of registers at each instruction, we perform
compiler analysis at the instruction level.
Determining the power state of each
register requires knowing the life time of registers as well as
the distance between the consecutive accesses to the
registers. We use the following notations.

\begin{itemize}
\item \IN{S} denotes the program point before the instruction $S$. 
       \OUT{S} denotes the program point after
  the instruction $S$.
\item \Succ{S} denotes the set of successors of the instruction $S$. An instruction $I$ is said to be successor of $S$ if the control may transfer to $I$ after executing the instruction $S$.
\item \islive{\pi}{R} is true if there is some path from program point $\pi$ to \fgexit\ that contains a use of $R$ not preceded by its definition.
\item \dist{\pi}{R} denotes the distance in terms of number of instructions from program point $\pi$ till the next access to $R$. \dist{\pi}{R} is set to $\infty$ when it exceeds the threshold $W$.
\item \sleepoff{\pi}{R} is true if the register $R$ can be put into SLEEP or OFF state at $\pi$. 
\item \power{\pi}{R} denotes the power state of the register $R$ at program point $\pi$. 
\end{itemize}

The liveness information of each register, \islive{\pi}{R}, can
be computed using traditional liveness
analysis~\cite{Khedker.DFA}. The data
flow equations to compute the \dist{\IN{S}}{R} and
\dist{\OUT{S}}{R} are as follows:
\[\begin{array}{rcl}
  \dist{\IN{S}}{R} &=& 
  \begin{cases}
    \mbox{1, if $S$ accesses $R$}\\
    \inc{\dist{\OUT{S}}{R}}, \mbox{otherwise}
\end{cases} \\
  \inc{x} &=&
  \begin{cases}
    \infty, \mbox{ if $x$ is $W$ or $\infty$} \\
    x + 1,  \mbox{otherwise} \\
  \end{cases}\\
  \dist{\OUT{S}}{R} &=& 
  \begin{cases}
    \infty, \mbox{ if $S$ is \fgexit\ }\\
    \max\limits_{SS \in \Succ{S}}\!\!\!\! \dist{\IN{SS}}{R},\ \mbox{otherwise}
  \end{cases}
\end{array}\]
Note that $\inc{x}$ is a saturating increment operator. Since our
analysis aims to reduce the power consumption, we compute
\dist{\OUT{S}}{R} as the maximum value of $\dist{\IN{SS}}{R}$ over
the successors $SS$ of $S$.  A register $R$ can potentially be put
into SLEEP or OFF state at a program point $\pi$ if it is not
accessed within the distance window $W$ on some path:
$$ \sleepoff{\pi}{R} = (\dist{\pi}{R}== \infty) $$

\begin{table}[t]
\centering
  \caption{Computing Power State of a Register $R$ at a Program Point $\pi$} \label{tab:power} \label{tab:power}
  \renewcommand{\arraystretch}{1.0}
  \begin{tabular}{ll|l}
    \hline
    \islive{\pi}{R} & \sleepoff{\pi}{R} & \power{\pi}{R}  \\ \hline  \hline
    true          & true           & SLEEP       \\ 
    true          & false          & ON          \\ 
    false         & true           & OFF         \\ 
    false         & false          & ON          \\
    \hline
  \end{tabular}
\end{table}

The power state of each register at each program point can
be computed according to Table~\ref{tab:power}. Note that in GPUs, 
all the threads of a warp execute the same instruction in SIMT manner, hence
power state computed by the analysis is applicable to 32 registers corresponding to the 32 threads of a warp.

\sethlcolor{lightgray}
\fboxrule=0.03pt
\newcommand{\modf}[1]{\fcolorbox{lightgray}{lightgray}{#1}}
\newcommand{\powstat}{\emph{Power\_State}\xspace}
\newcommand{\pslist}{\emph{Power\_State\_List}\xspace}

\subsection{Encoding Power States} \label{sec:encoding}

The power state (\powstat) of a register can be one of the three
states: \textbf{OFF}, \textbf{SLEEP}, or \textbf{ON}. Thus, it requires
two bits to represent \powstat of one register. Since the power
state can change after every instruction at run-time, we need to
encode the \powstat of the operand registers of an instruction in
the instruction itself.

PTXPlus instructions~\cite{GPGPUSIM} can support up to 4 source
and 4 destination registers. Encoding \powstat of all the
registers will require 16 bits. We observed that in our
benchmarks, most instructions use only up to 2 source registers
and 1 destination register. Therefore, to reduce the number of
bits required to encode \powstat in each instruction, we encode
information only for  2 source registers and 1 destination
register. For instructions having more registers, 
\powstat of the remaining registers is assumed to be \textbf{SLEEP} to
enable power saving. The modified instructions format is:
\vskip 1mm
\begin{center}
  \scalebox{.9}{\framebox[1.1\width]{{$<$Opcode$>$ $<$Options$>$ $<$Operand\_List$>$ \modf{$<$\pslist$>$}}}}
\vskip 3mm
\end{center}
where \power{\OUT{S}}{R} is \powstat encoded for a register $R$ for an instruction $S$.

\begin{figure}[t]
  \centering
  \begin{tabular}{@{\quad}c@{\quad\quad\quad}c}
    \scalebox{.75}{\tt\small
      \renewcommand{\arraystretch}{1.0}
      \begin{tabular}[b]{@{}r@{\ }|c@{}l@{}|} \cline{2-3}
1 & B4:& set.le.s32.s32 \$p2/\$o127, \$r8, \$r0, \modf{ON, SLEEP, ON}; \\
2 &&     ssy 0x00000110; \\
3 &&     mov.u32 \$r1, \$r0, \modf{SLEEP, ON}; \\
4 &&     \@\$p2.ne bra B8; \\ \cline{2-3}
5 & B5:& shl.u32 \$r10, \$r0, 0x00000002, \modf{ON, SLEEP}; \\
6 &&     mov.u32 \$r12, \$r124, \modf{ON, SLEEP}; \\
7 &&     add.half.u32 \$r11, s[0x0018], \$r10, \modf{ON, ON}; \\
8 &&     add.half.u32 \$r10, s[0x0020], \$r10, \modf{ON, ON}; \\ \cline{2-3}
9 & B6:& ld.global.u32 \$r14, [\$r11], \modf{ON}; \\
10 &&    ld.global.u32 \$r13, [\$r10], \modf{ON}; \\
11 &&    mad.f32 \$r12, \$r14, \$r13, \$r12, \modf{SLEEP, OFF, OFF}; \\
12 &&    add.u32 \$r1, \$r1, 0x00000400, \modf{ON, ON}; \\
13 &&    set.gt.s32.s32 \$p2/\$o127, \$r8, \$r1, \modf{ON, SLEEP, SLEEP}; \\
14 &&    add.u32 \$r10, \$r10, 0x00001000, \modf{SLEEP, SLEEP}; \\
15 &&    add.u32 \$r11, \$r11, 0x00001000, \modf{SLEEP, SLEEP}; \\
16 &&    \@\$p2.ne bra B6; \\ \cline{2-3}
17& B7:& bra B9; \\ \cline{2-3}
18& B8:& mov.u32 \$r12, \$r124, \modf{ON, SLEEP}; \\ \cline{2-3}
19& B9:& add.u32 \$r0, \$r0, \$r5, \modf{ON, ON, SLEEP}; \\
20 &&    shl.b32 \$ofs1, \$r9, 0x0, \modf{ON, ON}; \\
21 &&    set.le.s32.s32 \$p2/\$o127, \$r0, \$r6, \modf{ON, SLEEP, SLEEP}; \\
22 &&    mov.u32 s[\$ofs1+0x0000], \$r12, \modf{OFF}; \\
23 &&    add.u32 \$r9, \$r9, \$r7, \modf{SLEEP, SLEEP, SLEEP}; \\
24 &&    \@\$p2.ne bra B4; \\ \cline{2-3}
    \end{tabular}}
    &    
    \includegraphics[scale=0.5]{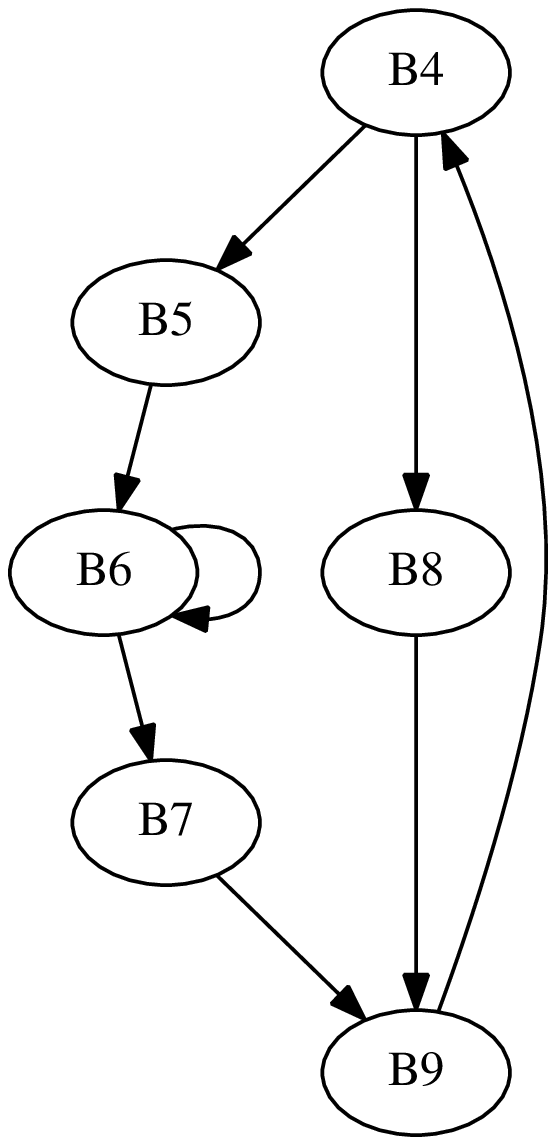}
    \\
    {\footnotesize (a) Power Optimized PTXPlus} & {\footnotesize$\qquad$ (b) CFG}
  \end{tabular}
  {\vskip 1mm \scriptsize The shaded text in part (a) denotes the power states inserted by \ourtool.
  }
 \vskip -2mm
\caption{A Snippet of the Program and its CFG for \emph{SP} Benchmark~\cite{CUDA-SDK}}
\label{fig:OptPtxpluls}

\vskip-3mm
\end{figure}

\begin{example}\label{ex:transformation}
Figure~\ref{fig:OptPtxpluls}(a) shows a snippet of power
optimized PTXPlus code, which is generated for \emph{SP}
benchmark using a threshold value ($W$) 7. The control flow graph (CFG) 
corresponding to the snippet is shown in Figure~\ref{fig:OptPtxpluls}(b). 
Note that the CFG is shown with respect to traditional basic block level to 
show it in compact. In Figure~\ref{fig:OptPtxpluls}(a), explicit branch addresses 
have been replaced by block labels for ease of understanding. 
The instruction  at Line-1 uses 2 source registers (r8, r0)
and 2 destination registers (p2, o127).
As discussed, our analysis inserts the power states only for 2
source registers and 1 destination register. In this case, the
power states \textbf{ON, SLEEP, ON} correspond to the registers p2,
r8, and r0 respectively. The power state of o127 register (the
fourth register in the instruction) is set to SLEEP state after
accessing the register.

For register r0 of the instruction,
the next access to the register occurs at Line-3 (at distance 2,
less than the threshold value 7).  Hence, the compiler inserts the power
state as ON. Register p2 is also kept in ON state for a similar reason.
For register r8 of the same instruction, the next access occurs along
two paths. One of the paths has a use at a distance of 8 (along
B5 at Line-13, $>$ 7), and the other has a definition after B9
(not shown in the figure). \ourtool keeps the register in SLEEP
state since there is a path along which the next access happens
after a distance $>$ 7.

Finally, consider register r13 accessed by the instruction at 
Line-11. There is no further access of r13 along any path in
the program. Therefore, the power state of r13 is set to OFF to save
power.  \hfill\qed
\end{example} 

At run-time, power state of the source registers are set after
the register contents have been read, i.e., in the read operands
phase in the GPU pipeline, and the power state of the destination
registers are set after the register contents have been written,
i.e., in the write back stage of the pipeline. The details of the
hardware implementation are discussed in
Section~\ref{sec:hardware}.

\newcommand{\InstList}{{\bf InstList}}
\newcommand{\ins}{{\bf ins}}
\newcommand{\I}{{\bf I}}
\newcommand{\R}{{\bf R}}
\newcommand{\W}{{\bf W}}
	
 
\subsection{Run-time Optimization} \label{sec:runtime}

Recall that the compiler analysis described in
Section~\ref{sec:companalysis} computes \dist{\OUT{S}}{R} as the
maximum distance value over all successors when \OUT{S} is a
branch point. This decision increases the chances of power
savings, but it can be suboptimal at run-time as shown by
the following example.

\begin{figure}[t]
\centering
\begin{subfigure}{.51\textwidth}
  \centering
  \includegraphics[scale=1.0]{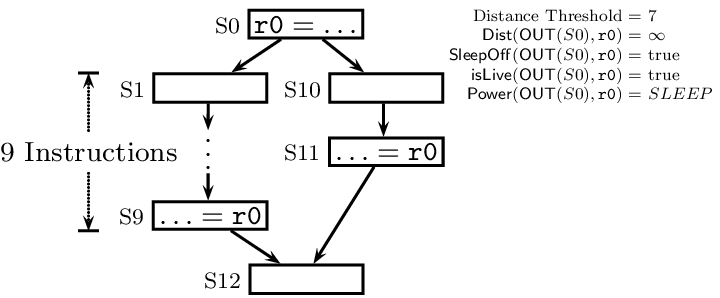} \vskip 0mm
  \caption{{\scriptsize Computing Distance at Branch Divergence}}
  \label{fig:sub1}
\end{subfigure}%
\begin{subfigure}{.49\textwidth}
  \vskip 5mm
  \centering\rule{0mm}{15mm}
  \includegraphics[scale=0.65]{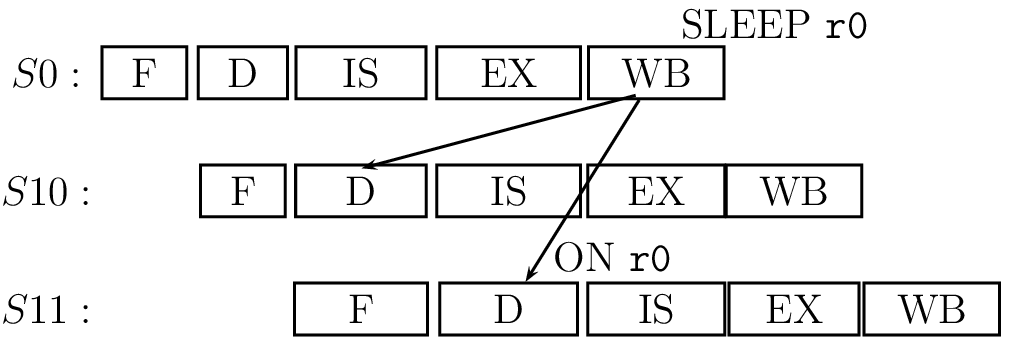}
  \caption{{\scriptsize Correcting Power State at Run-time.} }
\end{subfigure}

{\vskip 2mm \scriptsize The pipeline phases are:  Fetch (F), Decode (D), Issue (IS), Execute (EX), and Writeback (WB) }
\caption{Example for Run-time Optimization} 
\label{fig:runtime}
\vskip -2mm
\end{figure}

\begin{example} \label{ex:distance}
Consider the CFG in  Figure~\ref{fig:runtime}(a) for a hypothetical benchmark. Assume the
threshold value of 7 for \ourtool. Instruction S0 defines a
register r0. The next access to r0 occurs along two paths: the
path along S10 has a use at a distance of 2, and the other (along
S1) has a use in S9 at a distance of $\infty$ ($>$7).  
\ourtool computes \dist{\OUT{S0}}{r0} as $\infty$, the maximum of the
distances along the successors. Further, the state
\power{\OUT{S0}}{r0} is computed as SLEEP.  When the program
executes the path along S1, power is saved. However, if the
program executes the path along S10, then the register needs an
immediate wake up, causing an overhead.
\qed
\end{example}

\ourtool's compile-time decision can be corrected at 
run-time by looking at near future accesses of a register in the
pipeline. The hardware is modified to check in the 
pipeline if any decoded instruction from the same warp accesses a register
whose power state is being changed to SLEEP or OFF. 
If so, then the register power is
kept ON.  This avoids the wake up
latencies for instructions that access the same register within a
short duration, thereby avoiding the performance penalty.
Section~\ref{sec:hardware} describes the hardware implementation of
the optimization.

\begin{example} \label{ex:runtime}
Figure~\ref{fig:runtime}(b) shows a possible execution sequence of a
program whose CFG is shown in Figure~\ref{fig:runtime}(a). The
instruction $S0$ writes to
register r0.  After writing the register value in write back
stage (WB), the register needs to be put into SLEEP state. Assume
that the program takes the path along S10 and decodes the
instruction $S11$ before the write back stage of $  S0$. 
Our run-time optimization detects the future access to r0
by $S11$, and keeps the register in ON state instead of
putting it into SLEEP state to avoid additional wake up
latencies. On the other hand, if the program takes the path along
S1, then the instruction present in the S9 would appear much
later in the pipeline (after WB stage of $S0$). The
register r0 will be set to SLEEP state.  \hfill\qed
\end{example}

\begin{figure}[t]
\centering
  \includegraphics[scale=0.6]{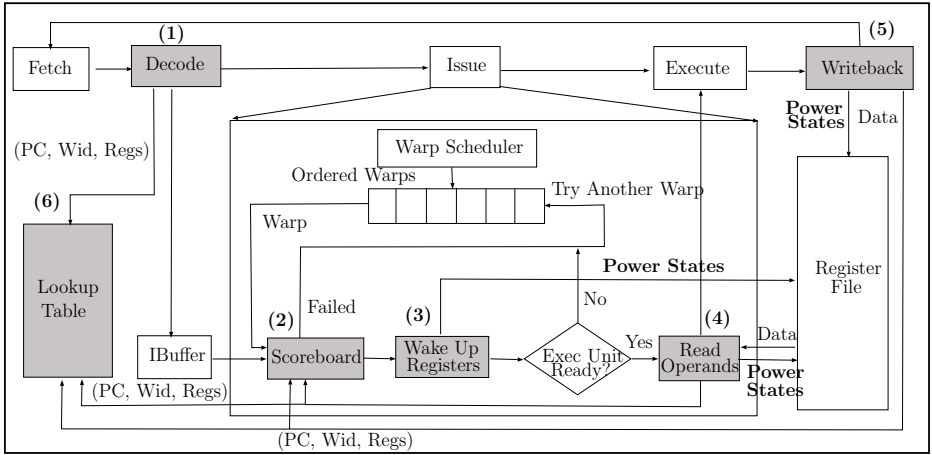} \vskip -1mm
  \caption{Modifications to GPU Pipeline}\label{fig:Architecture}\vskip -3mm
\end{figure}

Note that the effectiveness of run-time optimization depends on the application behavior at the branch divergent points, i.e., it is more effective when the power status estimated by the compiler analysis is sub-optimal at the divergent points. However, our experiments  (discussed in Section~\ref{sec:optimizations}) show that the compiler optimization is much more effective than the run-time optimization for the benchmarks used in the experiments.


\subsection{Hardware Support} \label{sec:hardware}
Figure~\ref{fig:Architecture} shows the modified pipeline of GPU Architecture that supports our proposed ideas, with the modified components shaded and labeled. The changes are described below and the corresponding overheads are quantified in Section~\ref{sec:overheads}.
\begin{enumerate}
\item To support the new instruction format (Section~\ref{sec:encoding}), we modify the decode unit to extract the power states of the registers from the instruction (Label (1) in Figure~\ref{fig:Architecture}).
\item The scoreboard unit (Label (2)) is modified to track RAR (Read After Read) and WAR (Write after Read) dependencies in addition to RAW (Read After Write) and WAW (Write after Write) dependencies. This is done by adding instruction's source registers in the scoreboard table. It is because an instruction can change the power state of a register to SLEEP or OFF  after reading the registers. Hence, the subsequent instructions that read/write the same register need to wait until the power state is modified.
\item The registers in SLEEP or OFF state are woken up by sending a wake up signal to the register file (Label (3)). A warp is considered ready for issuing its current instruction only when all its operand registers are in ON state.
\item The read operands phase (Label (4)) is modified (a) to set the power state of source registers after they have been read and (b) to release the source registers of the instruction which were reserved by the scoreboard unit.
\item The write back stage (Label (5)) includes the logic to set the power state of the destination registers after the registers are written.
\item The run-time optimization is implemented by adding a lookup table (Label (6)) to keep track of the registers accessed by an instruction. For an instruction having program counter \emph{PC} and warp id \emph{Wid}, the lookup table is indexed by \emph{Wid}. When an instruction is decoded, the decode unit inserts the instruction's operand registers into the lookup table. When a warp (\emph{Wid}) needs to set the power state of a register ($R$) of an instruction (\emph{PC}) to SLEEP or OFF, it searches the lookup table for another instruction (a different PC) with the same \emph{Wid} and accessing $R$. If a match is found, then the power state of $R$ is kept ON, otherwise, it is changed. After an instruction completes its writeback stage, the corresponding entry is removed from the lookup table.
\end{enumerate}

Each entry for a warp in the look up table stores instruction's \emph{PC}, and its register numbers. The number of entries required for each  warp is determined by the pipeline depth, which can be large. However, in practice, the number of entries required per each warp is less, and experimentally we found that  the average number of entries per warp is less than 2. If an SM allows maximum $W$ resident warps,  stores $w$ entries per each warp, supports $r$ operand registers for each instruction, and allows maximum $R$ registers per each thread, then the size of look up table (in bits) is $W*w*(sizeof(PC)+(log_2(R)*r))$. 


%% file: Sections/07-EvaluationMethodology.tex
\section{Evaluation Methodology}\label{sec:experimentanalysis} 

We implemented the proposed hardware changes and compiler optimizations in GPGPU-Sim V3.x~\cite{GPGPUSIM}. The
modified instruction format is implemented by extending the PTXParser provided by GPGPU-Sim.
The GPGPU-Sim configuration used for the experiments is shown in Table~\ref{table:GPGPUArch}.
We used \linebreak GPUWattch~\cite{GPUWattch} to measure the power consumption of register file.

 \begin{wraptable}{r}{6.5cm}
 \moveup
\caption{GPGPU-Sim Configuration}
\vskip -4mm
\centering
\scalebox{0.75}{\renewcommand{\arraystretch}{1.0}
\begin{tabular}{@{}l|l@{\ }}
  \hline\hline
  Resource & Configuration \\
  \hline
  Architecture & NVIDIA Tesla K20x\\
  Number of SMs & 14 \\
  Shader Core Clock	& 732 MHz \\ 
  Technology Node & 22nm \\
  Register File Size per SM \space \space \space & 256KB  \\ 
  Number of Register Banks & 32 \\
  Max Number of TBs per SM & 16 \\
  Max Number of Threads per SM & 2048  \\
  Warp Scheduling & LRR \\
  Number of Schedulers per SM & 4 \\ \hline
\end{tabular}}
\label{table:GPGPUArch}
\end{wraptable}

Note that GPUWattch internally uses CACTI~\cite{Cacti} to measure the power
dissipation that does not support leakage power saving
mechanism. 
Therefore, we modified GPUWattch to use CACTI-P~\cite{Cactip} that
provides power gating technique, which can minimize the leakage
power by setting the SRAM cells into low power (SLEEP or OFF)
state. It uses minimum data retention voltage so that SRAM cells
can enter into SLEEP state without losing their data.  We chose
SRAM$_{vccmin}$ to be the default value (provided by CACTI-P
depending on the technology node, 22nm for this case). To put
SRAM cells in OFF state, we configured
SRAM$_{vccmin}$ to 0 V.  In GPUs, the registers are allocated to
each warp in a private manner.  To independently turn the power
state of warp registers into OFF, SLEEP, and ON states, we set
the granularity of the subarray size in CACTI-P~\cite{Cactip} to 1 warp register (a set of
32 registers). After running several experiments, 
we chose the threshold value ($W$) as 3, which achieves lowest energy
for maximum number of kernels.  We used the
latency to wake up a register from SLEEP to ON state to be 1
cycle as reported in ~\cite{Cactip}, and the latency to wake up a 
register from OFF to ON state be twice (i.e., 2 cycles)~\cite{Mcpat}, 
except for Section~\ref{subsec:latency} where we consider the effect of 
other values for the wake up latencies on performance and energy consumption.
We report these latency and energy 
overheads in Section~\ref{sec:overheads} and also include these
overheads throughout our results.

We evaluated \ourtool on several applications from the benchmark suites CUDA-SDK~\cite{CUDA-SDK},  GPGPU-SIM~\cite{GPGPU-Sim},  Parboil~\cite{parboil}, and Rodinia~\cite{rodinia}. Table~\ref{table:benchmarks} shows the list of applications and kernel that is simulated for each application. We compiled all the applications using CUDA-4.0\footnote{GPGPU-Sim does not support above CUDA 4.0.}. We measured the effectiveness of our approach using the following metrics: (1) Power, (2) Energy, (3) Simulation Cycles.

\begin{table}[t]
  \caption{Benchmarks Used for Evaluation\label{table:benchmarks}}
\vskip -3mm  
  \scalebox{0.7}{\renewcommand{\arraystretch}{0.95}
    \begin{tabular}{l@{\ }l@{\quad }l@{\quad }l@{\quad }l@{\ }|l@{\ }l@{\quad }l@{ }l@{\ }l@{\ }l@{\ }l@{}}
     \hline \hline
        Sr. No. & Benchmark & Application  & Notation & Kernel                          & Sr. No. & Benchmark & Application & Notation & Kernel             \\ \hline
        1    & RODINIA   & backprop     & BP       & bpnn\_adjust & 12   & GPGPU-SIM & MUM         & MUM      & mummergpuKernel  \\ 
                 &    &      &        & weights\_cuda &    &  &          &       &   \\ 
        2    & RODINIA   & bfs          & BFS1     & Kernel                          & 13   & GPGPU-SIM & NN          & NN1      & executeFirstLayer  \\ 
        3    & RODINIA   & bfs          & BFS2     & Kernel2                         & 14   & GPGPU-SIM & NN          & NN2      & executeSecondLayer \\ 
        4    & CUDA-SDK  & Blackscholes & BS       & BlackScholesGPU                 & 15   & GPGPU-SIM & NN          & NN3      & executeThirdLayer  \\ 
        5    & RODINIA   & lavaMD       & LMD      & kernel\_gpu\_cuda                 & 16   & GPGPU-SIM & NN          & NN4      & executeFourthLayer \\ 
        6    & GPGPU-SIM & LIB          & LIB      & Pathcalc\_Portfolio\_ & 17   & RODINIA   & pathfinder  & PF       & dynproc\_kernel    \\
             &    &      &        & KernelGPU &    &  &          &       &   \\        
        7    & GPGPU-SIM & LPS          & LPS      & GPU\_laplace3d                  & 18   & CUDA-SDK  & scalarProd  & SP       & scalarProdGPU      \\ 
        8    & CUDA-SDK  & MonteCarlo   & MC1      & inverseCNDKernel                & 19   & PARBOIL   & sgemm       & SGEMM    & mysgemmNT          \\ 
        9    & CUDA-SDK  & MonteCarlo   & MC2      & MonteCarloOne & 20   & PARBOIL   & spmv        & SPMV     & spmv\_jds          \\ 
             &    &      &        & BlockPerOption &    &  &          &       &   \\  
        10   & PARBOIL   & mri-q        & MR1      & ComputePhi          & 21   & CUDA-SDK  & vectorAdd   & VA       & VecAdd             \\ 
             &    &      &        & Mag\_GPU &    &  &          &       &   \\          
        11   & PARBOIL   & mri-q        & MR2      & ComputeQ\_GPU                   & ~    & ~         & ~           & ~        & ~                  \\

\hline
    \end{tabular}
    }
\vskip -2mm    
\end{table}

%% file: Sections/08-ExperimentalAnalysis.tex
\section{Experimental Results} \label{sec:results}
In the results, we compare \ourtool with  \emph{Basline} approach, which is the default approach that does not use any leakage power saving mechanisms. Also, we compare it with warped register file~\cite{WarpedRegFile} technique (denoted as \emph{Sleep-Reg}), which comes closest to our work. \emph{Sleep-Reg} optimizes the baseline approach by (1) turning OFF the unallocated registers and (2) turning the allocated registers into SLEEP state immediately after the registers are accessed.

\begin{figure}[t]
\centering
\includegraphics[scale=0.40]{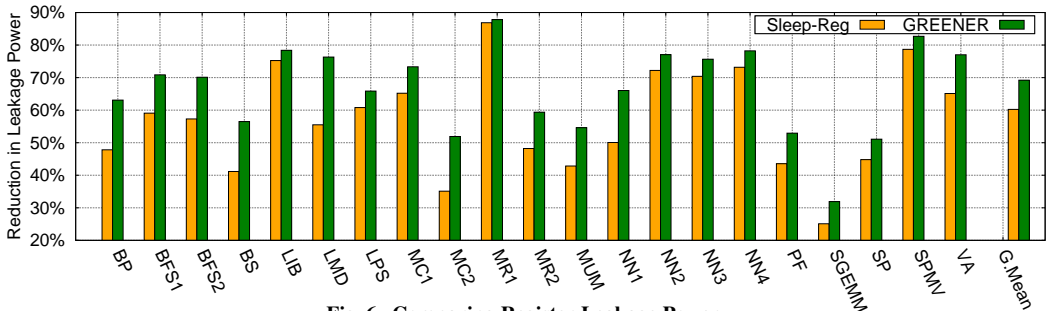}
\vskip -6mm
\caption{Comparing Register Leakage Power}
\label{fig:power}
\vskip -1mm
\end{figure}

\subsection{Comparing Register Leakage Power}
Figure~\ref{fig:power} shows the effectiveness of \ourtool\ and \emph{Sleep-Reg} by measuring the reduction in leakage power with respect to \emph{Baseline}. From the figure, we observe that \ourtool\ shows an average (Geometric Mean denoted as \emph{G.Mean}) reduction of leakage power by 69.21\% when compared to the \emph{Baseline}.  It shows the \ourtool\ is effective in turning the instruction registers into lower power state, such as SLEEP or OFF state depending on the behavior of the registers. The \emph{Baseline} does not provide any mechanism to save the leakage power, as a result, the registers of a warp continue to consume leakage power throughout the warp execution. Figure~\ref{fig:power} also shows that \emph{Sleep-Reg} approach reduces the register leakage power by 60.23\% when compared to \emph{Baseline}, however, \ourtool is more power efficient than \emph{Sleep-Reg}. It is because \emph{Sleep-Reg} approach reduces the leakage power by turning the instruction registers into SLEEP state immediately after the instruction operands are accessed, without considering the access pattern of the registers. If a register needs an immediate access, then keeping the register into SLEEP instead of ON state requires additional latency cycles to wake up the register, and 
during these additional cycles, the registers consume power. Further, \ourtool\ saves more leakage power compared to \emph{Sleep-Reg} by turning the registers into OFF state when there is no future use of the register, whereas \emph{Sleep-Reg} turns the register into only SLEEP state irrespective of its further usage.

\begin{figure}[t]
\centering
\includegraphics[scale=0.40]{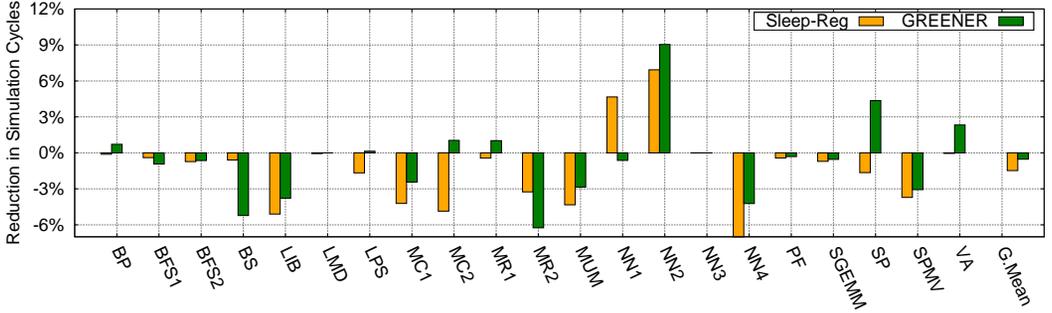}
\vskip -4mm
\caption{Comparing Performance in terms of Simulation Cycles}
\label{fig:performance}
\vskip -1mm
\end{figure}

\subsection{Performance Overhead Using Simulation Cycles} \label{sec:overhead}
Figure~\ref{fig:performance} shows the performance overheads of \ourtool\ and \emph{Sleep-Reg} approaches in terms of the number of simulation cycles with respect to \emph{Baseline}. 
On an average, the applications show a negligible performance overhead of 0.53\% with respect to \emph{Baseline}. A slowdown is expected because \ourtool\ turns the registers into SLEEP or OFF states to enable power savings, and these registers are turned back to ON state (woken up) when they need to be accessed. This wake up process takes few additional latency cycles which leads to increase in the number of simulation cycles. Interestingly, some applications (\emph{BP, LPS, MC2, MR1, NN2, SP}, and \emph{VA}) show improvement in their performance. This occurs due to the change in the issuing order of the instructions. The warps that require their registers to be woken up can not be issued in its current cycle,  instead other resident warps that are ready can be issued. This change in the issue order leads to change in the memory access patterns, which in turns changes L1 and L2 cache misses etc. In case of \emph{BP, LPS, MC2}, and \emph{NN1} applications, we observe an improvement in the performance due to less number of pipeline stall cycles with \ourtool\ when compared to \emph{Baseline}. \emph{MR1} shows less number of scoreboard stall cycles with \ourtool when compared to \emph{Baseline}. Though \emph{SP} and \emph{VA} applications have same number L1 and L2 cache misses with \ourtool\ and \emph{Baseline} approach, \ourtool\ shows less number of pipeline stall cycles when compared to \emph{Baseline}.

Figure~\ref{fig:performance} also shows that \emph{Sleep-Reg} has an average performance degradation of 1.48\% when compared to the \emph{Baseline} approach.  This degradation is more when compared to \ourtool\ because \emph{Sleep-Reg} turns all the instruction registers into SLEEP state after the instruction operands are accessed, irrespective of their usage pattern. If a register
in SLEEP state is accessed in near future, it needs to be turned on, this incurs additional wake up latencies with \emph{Sleep-Reg}. Whereas, our approach minimizes these additional wake up latency cycles by retaining such registers in the ON state. However, \emph{MR2} performs better with \emph{Sleep-Reg} because it shows less number of scoreboard and idle cycles than that of \ourtool.  Also, \emph{Sleep-Reg} performs better with \emph{BS} and \emph{NN1} since it has less number of stall cycles when compared to \ourtool.

\begin{figure}[t]
\centering
\includegraphics[scale=0.40]{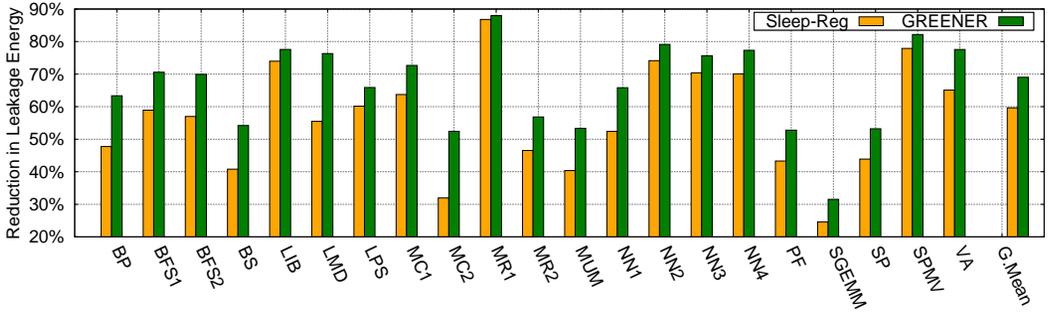}
\vskip -4mm
\caption{Comparing Register Leakage Energy}
\label{fig:energy}
\vskip -1mm
\end{figure}

\begin{figure*}
\centering
\includegraphics[scale=0.36]{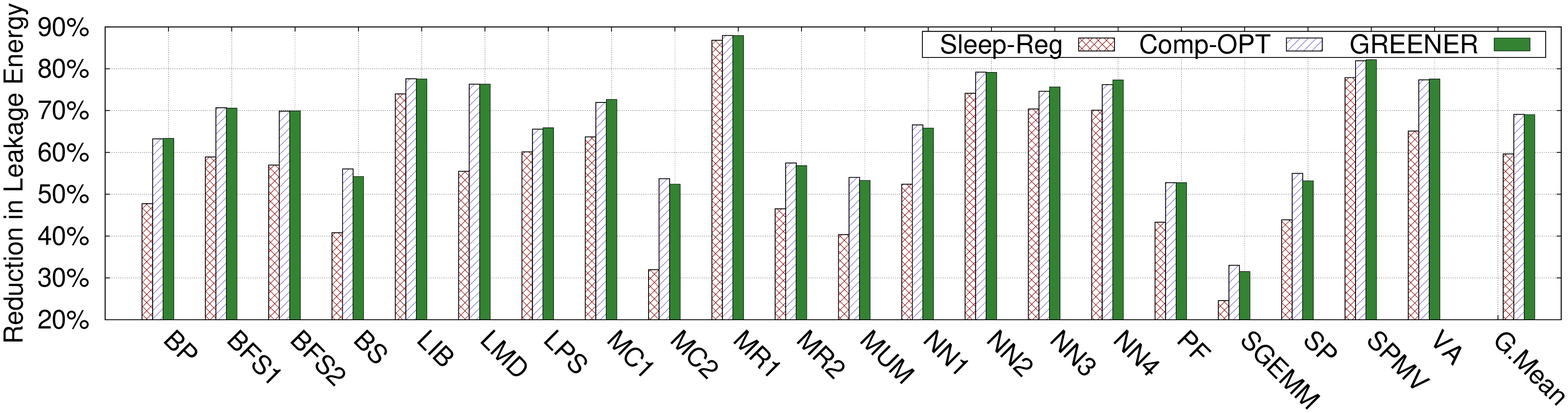}
\vskip -4mm
\caption{Comparing Effectiveness of Individual Optimizations}
\label{fig:optimizations}
\vskip -1mm
\end{figure*}

\subsection{Comparing Register Leakage Energy}

Figure~\ref{fig:energy} compares the total energy savings of \ourtool\ and \emph{Sleep-Reg} w.r.t. \emph{Baseline}. The results show that \ourtool\ achieves an average reduction of register leakage energy by 69.04\% and 23.29\% when compared to \emph{Baseline} and \emph{Sleep-Reg} respectively.  From Figures~\ref{fig:power} and \ref{fig:performance}, we see that \ourtool\ shows more leakage power saving, also has negligible performance overhead with respect to the \emph{Baseline}, hence we achieve a significant reduction in leakage energy. Also, the applications that exhibit more power savings and improve their performance with \ourtool, further show more leakage energy savings. Similarly, the applications that show leakage power savings but has more performance overhead will reduce their leakage energy savings accordingly when compared to \emph{Baseline} and \emph{Sleep-Reg} approaches.

\subsection{Effectiveness of Optimizations} \label{sec:optimizations}
We show the effectiveness of the proposed optimizations in Figure~\ref{fig:optimizations}.  From the figure, we analyze that the compiler optimization (discussed in Section~\ref{sec:companalysis}, and denoted as \emph{Comp-OPT}) saves more energy (average 69.09\%) when compared to \emph{Sleep-Reg} (59.65\%). This shows that turning the registers into low power states (SLEEP or OFF state) with the knowledge of register access pattern is more effective than turning the registers into SLEEP state after accessing them. 


The run-time optimization (discussed in Section~\ref{sec:runtime}) is evaluated by combining it with \emph{Comp-OPT}, and we denote them as  \ourtool\ in the figure. From the results, we observe that, for most of the applications,  \ourtool\ show minor improvements when compared to \emph{Comp-OPT} respectively. This is because the run-time optimization helps only in correcting power state of a register by turning to ON state when it detects the future access to the register at run-time. However, if the register is not found to be accessed in the near future at run-time, it does not modify and retains the power state  as directed by the \emph{Comp-OPT}. For some applications (e.g. \emph{NN3}),
\ourtool\ is less efficient when compared to \emph{Comp-OPT}. It occurs when a register that is determined to be accessed in the near future does not get accessed due to reasons such as scheduling order, scoreboard stalls, or the unavailability of the corresponding execution unit. In those cases, keeping the register into low power states (SLEEP or OFF) can save more energy instead of keeping it in ON state. Note that the effectiveness of run-time optimization depends on the application behavior at the branch divergence points.

\begin{figure*}
\centering
\includegraphics[scale=0.36]{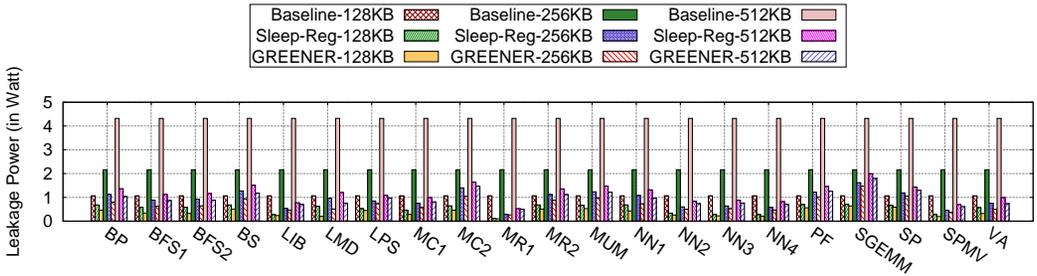}
\vskip -1mm
\caption{Comparing Leakage Power for Various Register File Sizes }
\label{fig:VaryRegisterFiles}
\vskip -1mm
\end{figure*}

\subsection{Leakage Power with Different Register File Sizes}

Figure~\ref{fig:VaryRegisterFiles} shows the effect of register
file size on leakage power for \emph{Baseline}, \emph{Sleep-Reg},
and \ourtool approaches. The register file sizes used are 128KB,
256KB, and 512KB. In the figure, \emph{Baseline-128KB} denotes
the \emph{Baseline} approach that is evaluated with 128KB
register file size. The other approaches are tagged in a similar way.

From the figure, we can see that for all the three approaches, leakage power increases with the increase in the register file size. This is because each memory cell in the register file consumes some amount of leakage power, and with the increase in the number of registers, leakage power increases. 
However, for \emph{Sleep-Reg} and \ourtool, this increase in the leakage power is less when compared to \emph{Baseline} since the register files in \emph{Baseline} consume leakage power irrespective of the register access. Whereas, registers can be put to lower power states with \emph{Sleep-Reg} and \ourtool. The results also indicate that \ourtool\ is effective when compared to \emph{Sleep-Reg} and \emph{Baseline} even with the increase in register file size. 

Interestingly, \ourtool\ with 512KB register file configuration consumes less amount of leakage power than that of \emph{Baseline} with 256KB configuration. Also, the leakage power of \ourtool\ for 512KB configuration is comparable to that of \emph{Baseline-128KB}. This shows that \ourtool\ with twice the register file size compared to \emph{Baseline}, not only shows less amount of leakage power consumption, but also can improve the amount of thread level parallelism.

\subsection{Analyzing Hardware Overheads} \label{sec:overheads}
To support leakage power saving, CACTI-P~\cite{Cactip} introduces additional sleep transistors into the SRAM structures. These transistors enable us to put the registers into low power states (SLEEP or OFF) after accessing the operands (discussed in Section~\ref{sec:hardware}), also they enable us to  wake up the registers from lower power states before accessing the operands. 
For the configuration used in our experiments, Table~\ref{table:measure_overhead} shows the additional area, latency, and energy associated with the additional sleep transistors circuitry. Note that in our experiments, we  conservatively consider the latency overhead to change the  power state of a register from OFF to ON state to be 2 cycles.

Recall that \ourtool encodes the power state of a register with its instruction, and we require 6 bits to encode the power states of the instruction registers. Currently, NVIDIA does not disclose the machine code format of the instructions. However, we can adopt either of the following two solutions as described in~\cite{VLIW}. (1) If the instruction format has 6 unused bits, we can exploit these bits to encode the power states. In this case, the instruction length would not increase, and there is no additional power overhead. (2) If there are no unused bits in the instruction format, we can extend the instruction length by 6 bits to encode the power states. However, this incurs additional storage in the GPU pipeline, such as instruction buffers overhead. We measure the additional overhead using GPUWattch framework by increasing the instruction length by 8 bits (2-bit padding for byte alignment). We observe that adding 8 bits to the instruction has $<$ 0.0001\% area overhead and $<$ 0.005\% leakage power overhead in each SM.

\begin{table}[t]
\caption{Hardware Overheads for Sleep Transistor Circuitry }
\centering
\scalebox{0.8}{\renewcommand{\arraystretch}{1.0}
\begin{tabular}{@{}l|l@{\ }}
  \hline\hline
  Parameter & Overhead \\
  \hline
  Area &  0.00875 $mm^2$\\
  Wake up Latency (SLEEP to ON) & 0.0197 $ns$ ($<$ 1 clock cycle) \\
  Wake up Latency (OFF to ON state) & 0.0551 $ns$ ($<$ 1 clock cycle) \\
  Energy (SLEEP to ON and vice versa) & 0.0633 $nJ$ \\ 
   Energy (OFF to ON and vice versa) & 0.198 $nJ$ \\  
  \hline
\end{tabular}}
\label{table:measure_overhead}
\end{table}

As discussed in Section~\ref{sec:hardware}, we are required to modify scoreboard unit in the scheduler unit to keep track of the read after read dependencies. Currently, GPUWattch does not support a power model for scoreboard unit. However, depending on the following design choices we may require additional overheads. (1) If the power model for scoreboard uses a bit mask to keep track of the registers accessed by a warp, then we do not require any additional storage overhead. We can use the existing bit mask to set the registers that will be read by a warp. (2) Instead, if the scoreboard explicitly maintains the register numbers accessed by each warp, then we need to store up to 4 source register numbers of an instruction. If each SM allows $W$ resident warps, and has $R$ registers per each thread, then the additional storage overhead for this scheme is $4*W*log_2(R)$. For the configuration used in our experiments (i.e., W=64, R=64), the storage overhead is 192 bytes, which is $<$ 0.1\% of register file size. Similarly, to support the run-time optimization, we require a look up table. For our experiments, the additional storage required for lookup table is 1280 bytes ($<$ 1\% of the register file size).

\begin{figure*}
\centering
\includegraphics[scale=0.36]{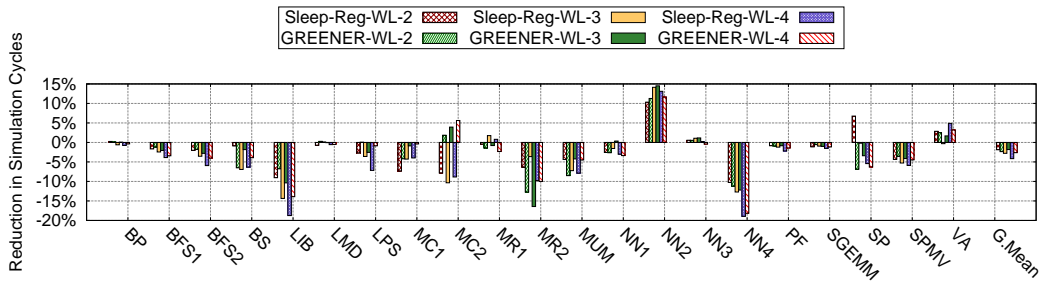}
\vskip -2mm
\caption{Comparing Performance Overhead for Various Wake Up Latencies}   
\label{fig:WakeupLatencies_Perf}     
\vskip -2mm
\end{figure*}

\begin{figure}[t]
\centering
\includegraphics[scale=0.36]{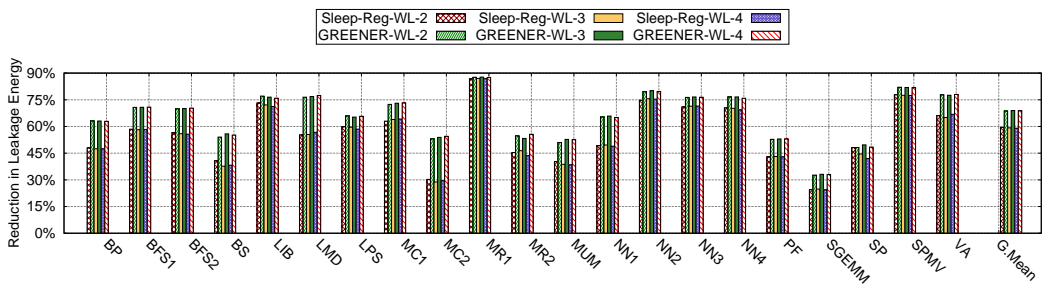}
\caption{Comparing the Leakage Energy for Various Wake Up Latencies}   
\label{fig:WakeupLatencies_energy}  
\end{figure}

\subsection{Effect of Wake up Latency} \label{subsec:latency}

Figure~\ref{fig:WakeupLatencies_Perf} compares performance overhead of \ourtool\ and \emph{Sleep-Reg} with the \emph{Baseline} for different values of wake up latencies.
In the figure, \ourtool-WL-X (X $\in$ \{2,3,4\})  denotes the \ourtool\ approach, which considers the  wake up latency to change a register state from SLEEP to ON to be X cycles. Whereas, when a register state needs to be changed from OFF to ON, it considers the latency to be 2X cycles. We use the similar notation for \emph{Sleep-Reg} as well.

For most of the applications, \ourtool\ and \emph{Sleep-Reg} show performance degradation with the increase in the wake up latency. 
The increase in the overhead is expected because applications spend additional simulation cycles for changing register's state from OFF or SLEEP state to ON state. Hence, with the increase in the wake up latency, these additional simulation cycles will increase. Interestingly, some applications (\emph{MC1 and MC2}) show performance improvement with the increase in the wake up latency. This is because, as discussed in Section~\ref{sec:overhead}, with the addition of wake up latency, the warps in the SM can get issued in different order, which can change the number of  L1-cache misses, L2-cache misses, and stall cycles etc. For MC1 and MC2, we find that the number pipeline stall cycles decrease with increase in the wake up latencies. Similarly, for \emph{NN2} we observe more number of L1 misses with \ourtool when used with wake latency 2 cycles than that of 3 cycles, hence \ourtool performs better with wake up latency 3 cycles.
Also, for \emph{NN2}, \ourtool\ performs better than \emph{Baseline} for all wake up latencies due to a decrease in the L1 misses when compared to \emph{Baseline}. In case of \emph{SP}, \ourtool-WL-2 has more stall cycles when compared to \ourtool-WL-3. Further, for most of the applications \ourtool\ performs better than \emph{Sleep-Reg} with various wake up latencies. 

We also compare the energy savings by varying the wake up latencies as shown in  Figure~\ref{fig:WakeupLatencies_energy}. The results indicate that even with varying the wake up latency, the applications show significant reduction in the leakage energy when compared to \emph{Baseline}. Also, the applications show more energy savings with \ourtool\ when compared to \emph{Sleep-Reg} for all wake up latencies.

\begin{figure}[t]
\centering
\includegraphics[scale=0.40]{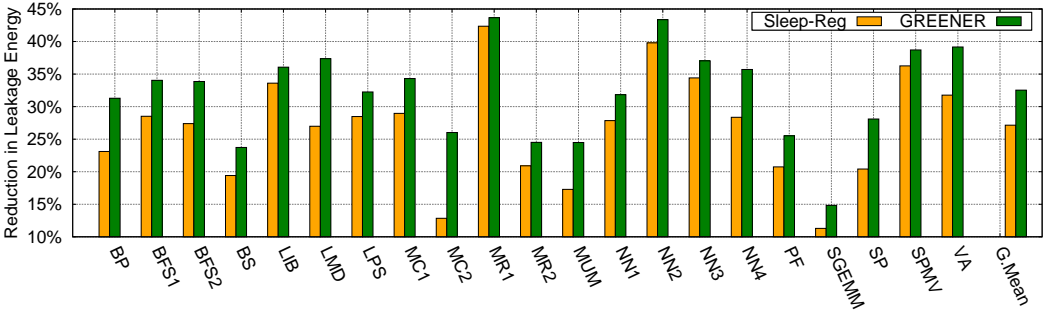}
\vskip -4mm
\caption{Comparing Leakage Energy by Including Routing Energy}
\label{fig:energy_with_routing}
\vskip -4mm
\end{figure}

\subsection{Leakage Energy Savings with Routing}
So far we discussed the energy efficiency of registers in a register file, however, 
GPUs also consume energy for routing of data and address through the register file. While modeling the register file, McPAT uses H-Tree distribution network to route data and address ~\cite{Mcpat}. The H-Tree network spends a constant amount of leakage power, and various organizations can be exploited to reduce this power and to meet routing requirement~\cite{Cacti}. Our work focuses only on reducing the leakage power of memory cells of the register file by analyzing the register access patterns, and reducing the routing power is not in this scope. However, we show the effectiveness of \ourtool\ by including the constant routing energy as shown in Figure~\ref{fig:energy_with_routing}.
From the results, we observe that \ourtool\ reduces the leakage energy on an average by 32.54\% when compared to \emph{Baseline}, which is more than that of \emph{Sleep-Reg} (27.15\%). However, the energy savings when including the routing energy are reduced when compared to that of results in Figure~\ref{fig:energy} because \ourtool\ does not provide any mechanism to optimize the routing power, hence the routing power remains unaffected.

\begin{figure}
\centering
\includegraphics[scale=0.4]{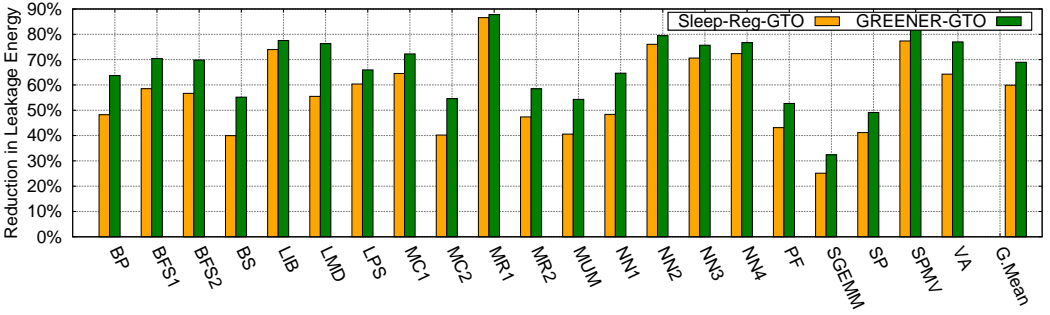}
\vskip -3mm
\caption{Comparing Leakage Energy using GTO Scheduler}
\label{fig:GTO}
\vskip -3mm
\end{figure}

\begin{figure}
\centering
\includegraphics[scale=0.4]{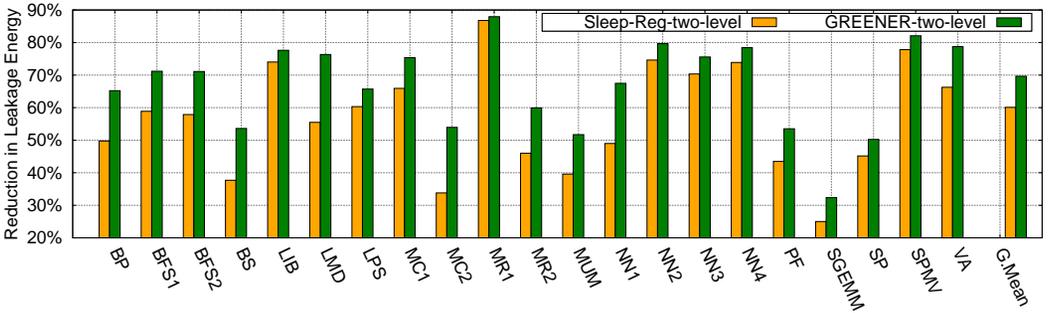}
\vskip -3mm
\caption{Comparing Leakage Energy using two-level Scheduler}
\label{fig:2Level}
\vskip -3mm
\end{figure}

\subsection{Leakage Energy Savings with Different Schedulers}

Figure~\ref{fig:GTO} and \ref{fig:2Level} show the effectiveness of \ourtool when it is evaluated with GTO and two-level scheduling policies respectively. The figures  compare \ourtool and \emph{Sleep-Reg} with \emph{Baseline} by measuring the reduction in leakage energy for the corresponding scheduling policies. The results show that \ourtool-GTO and \ourtool-two-level achieve an average reduction leakage energy by 68.95\% and 69.64\% with respect to \emph{Baseline-GTO} and \emph{Baseline-two-level} respectively. With different scheduling policies, the warps in the SM have different interleaving patterns, which affect the distance between the two consecutive accesses to a register. Even with the change in these access patterns, \ourtool shows reduction in leakage energy when compared to \emph{Baseline} and \emph{Sleep-Reg}.  We also find that \emph{Baseline-GTO} performs better than \emph{Baseline-two-level} in terms of simulation cycles, hence \emph{Baseline-GTO} relatively consumes less leakage energy when compared to \emph{Baseline-two-level}. However, the average energy savings of \ourtool are not affected significantly even with change in the scheduler.

\begin{figure*}
\centering
\includegraphics[scale=0.36]{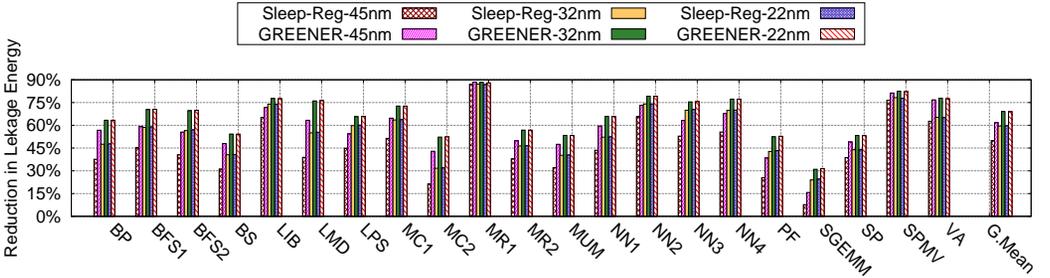}
\vskip -2mm
\caption{Comparing the Leakage Energy for Different Technology Configurations}
\label{fig:Technology}
\vskip -2mm
\end{figure*}

\subsection{Leakage Energy with Various Technologies }

Figure~\ref{fig:Technology} shows the effectiveness of \ourtool\ for various technology parameters (45nm, 32nm, and 22nm). The results show a significant reduction in leakage energy with \ourtool\ for all the applications even for various technology nodes. Further, it reduces the energy when compared to \emph{Sleep-Reg}. With transition of technology from 45nm to 32nm, we observe an increase in the leakage energy for the \emph{Baseline} approach, but \ourtool\ shows an increase in the leakage energy savings even with the transition. To model 22nm technology node, McPAT uses double gated technology to reduce the amount of leakage power, even with the advances in technology, \ourtool\ shows a reduction in leakage power when compared to \emph{Baseline}. To summarize, architectural techniques help in reducing the leakage power of a register file, in addition, the knowledge of register access patterns and compiler optimizations further help in reducing the leakage power and energy.

%% file: Sections/09-RelatedWork.tex
\section{Related Work} \label{sec:relatedwork}

Leakage and dynamic power are the two major sources of power
dissipation in CMOS technology. 
Reducing the leakage and dynamic power has been well studied
in the context of CPUs when compared to GPUs. 
Though \ourtool is only 
for saving leakage power consumption of GPU register files, we describe
briefly the techniques to save leakage and dynamic power in the
context of both CPUs as well as GPUs.
A comprehensive list of  architectural techniques to reduce 
leakage and dynamic power of
CPUs are described in~\cite{PowerTechniques}. Mittal et
al.~\cite{SurveryCPU} discuss the state of the art approaches for
reducing the power consumption of CPU register file. A survey of
methods to~\mbox{reduce} GPU power is presented in
\cite{SurveryGPU}.

\subsection{CPU Leakage Power Saving Techniques}
Powell et al.~\cite{Gatedvdd} proposed a state destroying
technique, Gated-$V_{dd}$, to minimize the leakage power of SRAM
cells by gating supply voltage. Several methods~\cite{DRI, decay,
  FineGrain} leverage Gated-$V_{dd}$ technique to reduce the
leakage power of cache memory by turning off the inactive cache
lines. However, these techniques cannot preserve the state of the
cache lines.  To maintain the state, Flautner et
al.~\cite{Drowsy} proposed an architectural technique that
reduces the leakage power by putting the cache lines into a
drowsy state.  Other approaches ~\cite{hotspot,drowsyimprv}
exploit this by using cache access patterns to put cache lines
in the drowsy state. As expected, the leakage power savings in this
(drowsy) approach are less when compared to Gated-$V_{dd}$
approach.

\subsection{GPU Leakage Power Saving Techniques}
Warped register file~\cite{WarpedRegFile} reduces leakage power
of register files by putting the registers into the drowsy state
immediately after accessing them. However, it does not take into
account the register access pattern while turning the registers
into low power states, hence it can have high overhead whenever there
are frequent wake up signals to the drowsy registers. In contrast,
\ourtool considers register access information and proposes
compile-time and run-time optimizations to make the register file
energy efficient. Their approach is closest to \ourtool and has been 
quantitatively compared in Section~\ref{sec:results}.

Register file virtualization~\cite{RegVirtual} reduces the
register leakage power by reallocating unused registers to
another warp. This uses additional meta instructions to turn off
the unused registers. However, the meta instructions are inserted
at every 18 instructions, which can cause a delay in turning off
the registers. \ourtool encodes the power saving states of the
registers in the same instruction, and hence the registers can be
switched to low power state at the earliest. Their approach
optimizes power for unused registers only, while \ourtool can put
even a used register into low power state if the next use is far
away in the execution.

Pilot register file~\cite{pilot} partitions the register file
into fast and slow register files, and it allocates the registers
into these parts depending on the frequency of the register
usage. It uses compiler and profiling information to allocate the
register into one of these parts.  The partition of the registers
is done statically. Therefore, if a register is accessed more
frequently for some duration, and less frequently for other
duration, then allocating the register to either of the
partitions can make it less energy efficient. \ourtool
changes power state during the execution, so it does not suffer
from this drawback.

Warped Gates~\cite{WarpedGates} exploits the idle execution units
to reduce the leakage power with a gating aware scheduling
policy. This approach is complementary to \ourtool and it
  should be possible to combine the two techniques to further reduce
  leakage power.

\subsection{Dynamic Power Saving Techniques for CPU and GPU}
In CPUs, dynamic voltage frequency scaling (DVFS) has been widely
adopted at the system level~\cite{SchedulingforEnergy}, compiler level~\cite{CompilerDVFS,CompilerForEnergy}, and hardware level~\cite{DVFSMCD}
to reduce dynamic power consumption.
In case of GPUs, equalizer~\cite{Equalizer} dynamically adjusts
the core and memory frequencies depending on the application
behavior and the user requirement (i.e., power or
performance). Lee et al.~\cite{GPUDVFS} propose mechanisms to
dynamically adjust the voltage and frequency values to improve
the throughput of applications under the power
constraints. GPUWattch~\cite{GPUWattch} uses DVFS algorithm to
reduce the dynamic power by adjusting the processor frequency
depending on the number of stall cycles.
Warped compression~\cite{Compression} exploits the register value
similarity to reduce effective register file size
to minimize the dynamic as well as leakage power. 
Gebhart et al.~\cite{Hierarchical} propose two complementary 
techniques to reduce GPU energy. The hierarchical register file 
proposed by them reduces register file energy by replacing 
the single register file with a multilevel  hierarchical register file. 
Further, they design a multi level scheduler that partitions warps to active 
and pending warps and propose mechanisms to schedule these warps 
to achieve energy efficiency.
These techniques mainly focus on reducing
the dynamic power of GPUs and are orthogonal to our approach.

\subsection{Miscellaneous}
Seth et al.~\cite{algorithmforpower} present algorithmic strategies for insertion of processor idle instructions at various points in the program such that the overall energy is reduced. 
Sami et al.~\cite{VLIW} employ liveness analysis to reduce the register file power in VLIW embedded architectures. Their approach exploits operand forwarding paths to minimize the number of register accesses of short lived registers. Hence, their mechanism is beneficial to the applications that have large number of short living registers.

%% file: Sections/10-Conclusion.tex
\section{Conclusions and Future Work} \label{sec:conclusion}

This paper focuses on reducing the leakage power of the register file in GPUs. We discuss various opportunities to save leakage power of the registers by analyzing the access patterns of the registers. We propose a new assembly instruction format that supports the power states of instruction's registers. Further, we provide a compiler analysis that determines the power state of each register at each program point, also transforms an input assembly  to power optimized assembly code. To improve the effectiveness further, we introduce a run-time optimization that dynamically corrects the power states determined by the static analysis.

We implemented the proposed ideas in GPGPU-Sim simulator and evaluated them on several kernels from CUDASDK, GPGPU-SIM, Parboil, and Rodinia benchmark suites. We achieved an average reduction of leakage energy by 69.04\% and maximum reduction of 87.95\% with a negligible performance overhead when compared to baseline approach.

The register leakage power constitutes a part of the total leakage power. Similarly, other resources in the GPU such as shared memory, cache, and DRAM, dissipate leakage power during a kernel execution. In future, we plan to work on reducing the power consumption of the other GPU resources by analyzing the application behavior and the resource access patterns. 